\documentclass{epl}
\title{Charge transfer counting statistics revisited}
\author{A. Shelankov\inst{1,2} \and J. Rammer\inst{1}}
\institute{
\inst{1} Department Physics, Ume{\aa} University, 
901 87 Ume{\aa}, Sweden \\
\inst{2} A. F.  Ioffe Physico-Technical
 Institute, 194021 St.Petersburg, Russia
}
\begin{document}
\maketitle
\begin{abstract}
Charge transfer statistics of quantum particles is obtained by
analysing the time evolution of the many-body wave function.
Exploiting properly chosen gauge transformations, we construct the
probabilities for transfers of a discrete number of particles.
Generally, the derived formula for counting statistics differs from
the one previously obtained by Levitov {\it et al.} (J. of
Math. Phys. {\bf 37}, 4845 (1996)).  The two formulae agree only if the
initial state is prohibited from being a superposition of different
charge states.  Their difference is illustrated for cases of a single
particle and a tunnel junction, and the role of charge coherence is
demonstrated.
\end{abstract}

Recently the question of counting statistics of charge transfer has
attracted considerable interest due to its relevance to electronic
transport in nanostructures where the discreteness of the electronic
charge is reflected in quantum transport properties
\cite{LevLeeLes96,Les89,Khl87,But90,BeeBut92,BeeSch01}.  Inspired by
the concept of photon counting in optics, counting statistics of
particles addresses a fundamental question of quantum transport, viz.
the probability distribution for the number of charges transferred
between different spatial regions of a system in a given time span.
The objective is to get complete information about the fluctuations in
particle currents, {\it i.e.}, correlations of any order.  Generally,
cross-correlations in charge transfers in different conducting
channels of a mesoscopic system are of interest \cite{BorBelBru02}.
In a seminal paper by Levitov, Lee and Lesovik \cite{LevLeeLes96}, a
formula for counting statistics was proposed by considering a {\em
gedanken} experiment in which a spin is coupled to the electrons in a
quantum wire whose transfer statistics are to be counted. The
precession of the spin then counts the number of electrons passing
either to the left or to the right of a chosen point in the wire.
Applying quantum measurement arguments to a system interacting with an
idealized measuring device, counting statistics was also considered by
Nazarov and Kindermann \cite{NazKin}.

Here, we shall develop counting statistics from a different point of
view; instead of analysing a measurement process, we extract
information about particle transfer directly from the wave function of
a many-body system.  We derive a formula for counting statistics which
turns out to differ from the one of Ref.~\cite{LevLeeLes96}.  Our
approach allows us to establish the circumstances under which the
counting formula obtained in Ref.~\cite{LevLeeLes96} is not
applicable.

In classical mechanics, the notion of counting statistics is
unproblematic: when complete information about a system is known --
the trajectories of individual particles are known -- there is a
unique answer to the question of how many particles are transferred
from, say, the left to the right in a given time interval.  In quantum
mechanics the situation is not so innocent.  Even when full
information about a many-body system is available, i.e., its
time-dependent wave function is known, there is no straightforward
algorithm to extract the probabilities in question, since there is no
quantum operator representing the number of transferred particles.  To
circumvent this difficulty, a {\em gedanken} experiment was in Ref.
\cite{LevLeeLes96} used as the basis for establishing a counting
formula: In the experiment, the rotation $\chi (\lambda )$ of a spin
coupled to the charge current via a gauge field was ``measured'' as a
function of the coupling constant $\lambda $ (and measuring time
interval $\tau $).  {\it Assuming} the generating function $\chi
(\lambda )$ to be $2\pi$-periodic, the Fourier coefficients in front
of $\exp(i m \lambda)$ were interpreted as the probabilities for the
passage of $m$ particles.  We point out that the {\it interpretation}
of the experiment is based on ingenious intuition rather than
following unequivocally from the principles of quantum mechanics.
Besides, it contains an ambiguity: as we show later, $\chi (\lambda )$
may have $\exp(\pm i \lambda/2 )$ components, which could then be
interpreted as half-integer charge transfers.  Also, positive
definiteness of the prescribed probabilities cannot be established.  A
{\em gedanken} experiment, not being a realistic one, is in fact a
vehicle for analysing the wave function of a system. Therefore we
shall develop an approach to counting statistics of quantum particles
which is based solely on an analysis of the wave function.

We consider a system partitioned into two parts by the plane at $x=0$,
referred to as left and right.  We want to tag the particles with a
``non-demolishing marker,'' {\it i.e.}, a marker that does not disturb
the quantum dynamics. The marker should provide information on whether
a particle has crossed the interface and in what direction.  As
explained below, this can be achieved by introducing the gauge
transformation
\begin{equation}
\hat{U}_{\lambda }= \exp\left[i \lambda \sum\limits_{k}\theta (-x_{k})\right] ,
\label{vrc}
\end{equation}
where $x_{k}$ is the coordinate of the $k$-th particle of the
many-body system.  In order to demonstrate how the gauge
transformation serves as a marker, we consider the case of a single
particle subject to a potential.  Let $\psi_{L}(x)$ and $\psi_{R}(x)$
denote normalized initial wave packet states located only on the left
or right, respectively.  A state is evolved in time according to $\psi
( \tau )= {\cal E}\psi (0)$, where ${\cal E}= \exp[-i{\cal H} \tau ]$
and ${\cal H}$ is the Hamiltonian of the system.  For each initial
state, specified at time $t=0$, the time evolution operator ${\cal E}$
produces a state which is a coherent superposition of left and right
components: $${\cal E} \psi_{L} = \psi_{L \rightarrow L} + \psi_{L
\rightarrow R} \quad,\quad {\cal E} \psi_{R} = \psi_{R \rightarrow L}
+ \psi_{R \rightarrow {R}} \quad, $$ where the last symbol in the
subscript on the r.h.s.  indicates the location of the wave packet to
be on the left (L) or right (R).  Of importance will be the gauge
transformed evolution operator
\begin{equation}
{\cal E}_{\lambda }= U_{\lambda}^{\dagger}{\cal
E}U_{\lambda}. 
\label{czc}
\end{equation} 
Letting it operate on the considered initial states, the following
marked final states emerges $${\cal E}_{\lambda } \psi_{L} = \psi_{L
\rightarrow L} + e^{i\lambda } \, \psi_{L \rightarrow R} \quad,\quad
{\cal E}_{\lambda } \psi_{R} = e^{- i\lambda } \, \psi_{R \rightarrow
L} + \psi_{R \rightarrow {R}} \quad ,$$ indeed states exhibiting the
intended transfer marking.  One immediately realises that the weights
$||\psi_{L \rightarrow L} ||^2 = \langle \psi_{L\rightarrow L}|
\psi_{L\rightarrow L} \rangle$ and $||\psi_{L \rightarrow R} ||^2=
\langle \psi_{L\rightarrow R}| \psi_{L\rightarrow R} \rangle $ are the
probabilities for the charge transfers $m=0$ and $m=1$, respectively.
Analogously, $||\psi_{R \rightarrow L} ||^2$ and $||\psi_{R
\rightarrow R} ||^2$ are the probabilities for transfers $m=-1$ and
$m=0$ for the initial state on the right.  One is able to extract the
components, $\psi_{m}$, of the final states ${\cal E} \psi_{L} $ and $
{\cal E} \psi_{R} $ which corresponds to $m$ particle transfers from
left to right by the following operation
\begin{equation}
\psi_{m} = \int\limits_{0}^{2\pi } \frac{d \lambda }{2\pi }
\; e^{-i m \lambda }\;{\cal E}_{\lambda } \psi^{(0)} ,
\label{2rc}
\end{equation}
where $\psi^{(0)}$ is the initial state, i.e., either $\psi_{L}$ or
$\psi_{R}$. The probabilities for the possible particle transfers can
therefore also be expressed on the form $P_m =
\langle\psi_{m}|\psi_{m}\rangle$.

Next we demonstrate that the procedure works for an arbitrary initial
state and consider a superposition of the wave functions located on
the left and right, $\psi^{(0)} = A \psi_{L} + B \psi_{R} $.  The
operation in Eq.~(\ref{2rc}) then produces the three charge transfer
states
\[
\psi_{m} = \left\{
\begin{array}{rcc}
A\psi_{L \rightarrow R}   &\;\;,\;\;& m=1  \;;\\
A \psi_{L \rightarrow L} + B\psi_{R \rightarrow R}
&\;\;,\;\;& m=0  \;;\\
 B \psi_{R \rightarrow L}  &\;\;,\;\;& m=-1      \;.
\end{array}
\right.
\]
We note that the weight of these states, $|| \psi_{m}||^2 = \langle
\psi_{m}| \psi_{m}\rangle $, are the probabilities for transfers of
$m$ particles for a general initial state. Indeed,
$\langle\psi_{1}|\psi_{1}\rangle = |A|^2 \langle\psi_{L \rightarrow
R}|\psi_{L \rightarrow R}\rangle$ is the product of the probability
initially to be on the left side and the conditional probability to
transfer from the left to the right, and analogously for the terms
$\langle\psi_{0}|\psi_{0}\rangle$ and
$\langle\psi_{-1}|\psi_{-1}\rangle$.

Using Eq.~(\ref{2rc}), the transfer probability, $P_m =
\langle\psi_{m}|\psi_{m}\rangle$, can be expressed as
\begin{equation} 
P_{m} = 
\int\limits_{0}^{2\pi }\frac{d\Lambda}{2\pi}
\frac{d\lambda}{2\pi}e^{-i m \lambda}
\langle\psi^{(0)}| {\cal E}_{\Lambda - \frac{\lambda}{2}}^{\dagger} 
{\cal E}_{\Lambda+\frac{\lambda}{2}}|\psi^{(0)}\rangle ,
\label{5rc}
\end{equation}
and the generating function, $\chi_{\tau }(\lambda )=
\sum\limits_{m}P_{m} e^{i m \lambda }$, reads
\begin{equation}
\chi_{\tau}(\lambda) = \int\limits_{0}^{2\pi} \frac{d \Lambda }{2\pi } 
\; \chi_{\tau} (\lambda , \Lambda )\;\; , \;\;
\chi_{\tau} (\lambda, \Lambda  ) = 
\langle{\cal E}_{\Lambda - \frac{\lambda}{2}}^{\dagger} 
{\cal E}_{\Lambda+\frac{\lambda}{2}} \rangle\, ,
\label{9rc}
\end{equation}
where the averaging is with respect to the initial state $\psi^{(0)}$,
or the density matrix of the system.  This expression is valid for a
many-body system provided the corresponding gauge-transformed
evolution operator Eq.~(\ref{czc}) is used.  Inverting the argument,
Fourier transformation with respect to $\lambda $ of the $2\pi
-$periodic function $\chi_{\tau}(\lambda)$, Eq.~(\ref{9rc}), generates
the coefficients $P_{m}$ (for integer $m$'s) which are guaranteed to
be positive, and $\sum_{m}P_{m}=1$.  Their meaning is that of
probabilities for integer charge transfers.

Expressing the evolution operator via the Hamiltonian, the integrand
of the generating function in Eq.~(\ref{9rc}) can be written as
\begin{equation}
\chi_{\tau}(\lambda, \Lambda  ) =
\left\langle 
T_{K}\exp \left[- i  \int\limits_{{\cal C}_{\tau}} dt'\; 
{\cal H}_{\gamma (t')}(t')\right]
\right\rangle_{0} ,
\label{tsc}
\end{equation}  
where ${\cal H}_{\gamma } = U_{\gamma } {\cal H}
U_{\gamma}^{\dagger}$, and ${\cal H}$ is the Hamiltonian for the
system, and the Keldysh contour ${\cal C}_{\tau}$ proceeds from
$t_{-}=- \infty $ to $\tau $ and back again as $t_{+}$ from $\tau $ to
$- \infty $; $T_{K}$ denotes the time ordering on the contour. The
projecting gauge field $\gamma $ is zero outside the measuring
interval, and $\gamma (t'_{\mp}) = \Lambda \pm \frac{\lambda}{2} $ for
$0< t' < \tau $, and the average in Eq.~(\ref{tsc}) is taken with
respect to the density matrix in the far past. The expression in
Eq.~(\ref{tsc}) can be evaluated using standard field theoretical
methods.

The formula, Eq.~(\ref{9rc}), differs from the generating function
proposed by Levitov {\it et al.} \cite{LevLeeLes96,LevRez01}. The
latter, denoting it $\chi_{\tau}^{\rm{L}} $, is obtained by setting
$\Lambda =0$ in Eq.~(\ref{9rc})
\begin{equation}
\chi_{\tau}^{\rm{L}}(\lambda )= 
\langle
{\cal E}_{-\frac{\lambda}{2}}^{\dagger} {\cal
  E}_{\frac{\lambda}{2}}
\rangle .
\label{maK}
\end{equation}
To verify that the two counting formulas are not equivalent, we
calculate $\chi_{\tau}^{\rm{L}}(\lambda )$ for the single particle
case where the initial state is the previously considered
superposition of right and left located wave packets, and obtain
\begin{equation}
\chi_{\tau}^{\rm{L}} (\lambda )  =   \chi_{\tau }(\lambda ) +     
4i \sin \frac{\lambda}{2} \Re
\left(A^{*} B \, \langle 
\psi_{L\rightarrow R} | \psi_{R\rightarrow R} \rangle\right).
\label{esc5}
\end{equation}
Indeed, $\chi_{\tau }^{\rm{L}}$ differs from our generating function.
We find that the difference is the additional term in $\chi_{\tau
}^{\rm{L}}$ which is $4\pi -$periodic in $\lambda $.  In
Ref.~\cite{LevLeeLes96} where the procedure for charge transfer
counting were based on the Fourier expansion of
$\chi_{\tau}^{\rm{L}}(\lambda )$, this implies that half-integer
charge transfers would occur.

To investigate further the difference between the two approaches, we
consider counting statistics from a different perspective.
Introducing the Hermitian operators
\[
{\cal P}_{n} = \int\limits_{0}^{2\pi } \frac{d\gamma}{2\pi} e^{-i n
\gamma }U_{\gamma } 
\quad,\quad n = 0, \pm 1, \ldots ,
\label{fsc}
\]
through the marker gauge transformation, Eq.~(\ref{vrc}), we realize
their meaning by noting that the operator ${\cal P}_{n}$ projects a
state $|\psi \rangle $ onto the component $|\psi_{n}\rangle = {\cal
P}_{n}|\psi \rangle $ which corresponds to exactly $n$ particles on
the left.  These projection operators, similar to the ones introduced
by P.~W.~Anderson in superconductivity, turn out to be suitable tools
for the kind of vivisection of a quantum state needed to obtain the
probability distribution for discrete charge transfers.  Being states
with definite particle number on the left, the projections
$|\psi_{n}\rangle $ are eigenfunctions of the operator $U_{\gamma }$,
$ U_{\gamma } |\psi_{n}\rangle = e^{i \gamma n} |\psi_{n}\rangle
$. This property can be expressed on operator form as ${\cal P}_{n} =
e^{- i \gamma n} U_{\gamma }{\cal P}_{n} $, whereby $ \sum_{n} e^{i n
\gamma }{\cal P}_{n} = U_{\gamma } \,.  $ Consequently,
Eq.~(\ref{5rc}) can be transformed into
\begin{equation}
P_{m} = 
\sum_{n}
\langle\psi^{(0)}| 
{\cal P}_{n}{\cal E}^{\dagger} {\cal P}_{n-m}
{\cal E} {\cal P}_{n}
|\psi^{(0)}\rangle ,
\label{psc}
\end{equation}
producing a different way of expressing the transfer probability.
According to quantum mechanics, the matrix element $\langle\psi^{(0)}|
{\cal P}_{n}{\cal E}^{\dagger} {\cal P}_{n-m} {\cal E} {\cal P}_{n}
|\psi^{(0)}\rangle = ||{\cal P}_{n-m} {\cal E} {\cal P}_{n} \psi^{(0)}
||^2$ is the probability for the transition from a state with $n$
particles on the left to a state with $n-m$ particles on the left. The
quantity $P_{m}$ is thus the probability for a transfer of $m$
particles to the right in a time span $\tau$ {\em given} that a
measurement of the charge state is performed initially.  For an
arbitrary mixture of states,
\begin{equation}
P_{m} = \sum\limits_{n}
\;\langle
{\cal P}_{n}{\cal E}^{\dagger} {\cal P}_{n-m}
{\cal E} {\cal P}_{n}
\rangle
\; ,
\label{psc02}
\end{equation}
where, as in Eq.~(\ref{9rc}), the average means taking trace with
respect to the density matrix $\rho_{0}$ at time $t=0$ when the
counting is initiated.  (We recall that the evolution operator ${\cal
E}$ evolves the system from time $t=0$ to $t= \tau$.)  For a classical
statistical ensemble, this is how the statistics of particle transfers
is evaluated, and we conclude that Eq.~(\ref{psc02}) and therefore the
generating function, Eq.~(\ref{9rc}), indeed has the correct classical
limit.

In terms of the charge projection operators, the generating function
Eq.~(\ref{9rc}) becomes
\begin{equation}
\chi_{\tau}(\lambda ) = \sum\limits_{n}
\;
\langle
{\cal P}_{n}{\cal E}^{\dagger}_{-\frac{\lambda}{2}}
{\cal E}_{\frac{\lambda}{2}} {\cal P}_{n}
\rangle \, .
\label{xvc02}
\end{equation}
We infer from this expression that if $\rho_{0}$ is diagonal in the
representation of the charge states, one of the projection operators
can be removed from Eq.~(\ref{xvc02}), and the sum over the remaining
projectors is unity; equivalently, $\chi_{\tau }(\lambda, \Lambda )$
does not depend on $\Lambda $ and the integration with respect to
$\Lambda $ can be omitted.  In {\it this} case, the generating
function, Eq.~(\ref{xvc02}), reduces to the form on the r.h.s. of
Eq.~(\ref{maK}), which is identical to the result of
Ref.~\cite{LevLeeLes96}.  The physical origin of the difference
between the two formulae is thus the charge off-diagonal components of
the density matrix. As one can show, it is the latter which produces
the unphysical $4\pi $-periodic part of the generating function of
Ref.~\cite{LevLeeLes96}.

We observe that the simple physical picture in which the current is
build of transfers of integer number of particles meets with
difficulties for a general initial state, when the system may be in a
superposition of different charge states, and $\rho_{0}$ is
non-diagonal in charge space.  Indeed, one expects that $Q(\tau
)=e\sum_{m} m P_{m}$ should equal the average charge transfer,
$\int_{0}^{\tau } \langle \hat{I}(t) \rangle dt$, $\hat{I}$ being the
Heisenberg current operator. Calculating $\dot{Q}$ from
Eq.~(\ref{9rc}) or Eq.~(\ref{xvc02}), $ \dot{Q}(t) = \sum_{n} \langle
{\cal P}_{n} \hat{I}(t) {\cal P}_{n} \rangle \;.  \label{nwc} $
Clearly, $\dot{Q}$ is identical to $\langle \hat{I} \rangle $ only
when $\rho_{0}$ is charge diagonal. A similar difficulty emerges for
the generating function of Ref.\cite{LevLeeLes96}: the expression $Q=
\sum_{m} m P_{m}^{L}$ gives the correct expectation value for the
transferred charge {\em only} if one allows $m$ to assume {\em
half-integer} as well as integer values, the former being due to the
$4 \pi $-periodic part of $\chi_{\tau }^{L}(\lambda )$ generated by
the charge off-diagonal elements of $\rho_{0}$.  The analysis points
to an ambiguity in counting statistics, a trade-off between having a
probability distribution for discrete charge transfers and the
generation of proper current correlation functions. The resolution is
shown to be tied to the charge structure of the initial state of the
system.

As an illustration, we evaluate the counting statistics for a tunnel
junction using Eq.~(\ref{tsc}); the problem was considered previously
in Ref. \cite{LevRez01} using the generating function in
Eq.~(\ref{maK}).  The system consists of two weakly connected metallic
regions, left and right.  The Hamiltonian reads $ {\cal H} = {\cal
H}_{0} + V^{T}$, where ${\cal H}_{0}$ refers to the isolated regions
and $V^{T}= V_{r \leftarrow l} + V_{l \leftarrow r}$ is the tunneling
part, where $V_{r \leftarrow l}$ ( $V_{l \leftrightarrow r}$) describe
transitions from left to right (right to left), $V_{r \leftarrow l}=
V_{l \leftarrow r }^{\dagger} $.  The gauge transformation $U_{\gamma
}$ affects only the tunneling part, and in Eq.~(\ref{tsc}),
\begin{equation}
{\cal H}_{\gamma } = {\cal H}_{0} + V_{\gamma }^{T}\;,\;
 V_{\gamma }^{T}=
e^{i \gamma }V_{r \leftarrow l}+ e^{- i \gamma }V_{l \leftarrow r}\,.
\label{zsc}
\end{equation}
As in Ref. \cite{LevRez01}, we consider only the leading contributions
with respect to tunneling in Eq.~(\ref{tsc}), and $W(\lambda,\Lambda)
\equiv \ln \chi (\lambda , \Lambda )$ can be evaluated as
\begin{equation}
W(\lambda,\Lambda) = 
- \frac{1}{2}
 \left\langle T_{K}
\int\limits_{- \infty }^{\tau} 
dt_{1} dt_{2}
\hat{V}_{\gamma (t_{1})}^{T}(t_{1})
\hat{V}_{\gamma (t_{2})}^{T}(t_{2})
\right\rangle ,
\label{yvc}
\end{equation}
where $\hat{V}^{T}(t)$ is the tunneling operator in the interaction
picture; for given $\lambda $ and $\Lambda $, the projecting field
$\gamma $ is the function of the Keldysh time $t_{\pm}$ introduced
below Eq.~(\ref{tsc}).  One obtains two contributions, $W = W_{MM}+
W_{MP}$. The first term, $W_{MM}$, originates from the time domain
where both time arguments $t_{1}$ and $t_{2}$ in Eq.~(\ref{yvc})
belong to the measuring interval from 0 to $\tau $, and the second
one, $W_{MP}$, is the contribution from the region where one of the
times $t_{1}$ or $t_{2}$ is prior to the start of the measurement,
$t=0$.

The term $W_{MM}$ is $2\pi $-periodic in $\lambda $ and does not
depend on $\Lambda $. It coincides with the result of
Ref.~\cite{LevRez01}: $ W_{MM} = (e^{i \lambda} -1)\; w_{+}(\tau ) \;
+ \; (e^{- i \lambda} -1)\; w_{-}(\tau ) $ where
\[
w_{+}(\tau )
=
2\pi \tau
\int dE\, d \epsilon \,
T_{E_{-}, E_{+}}\,
n_{E_{-}}^{l} 
\left(1 - n_{E_{+}}^{r}\right)
\Delta_{\tau }(\epsilon)
\quad,\quad \Delta_{\tau }(\epsilon)
=
\frac{1}{2 \pi \tau}
\left|
\frac{e^{i \epsilon \tau} - 1}{\epsilon}
\right|^2 \; .
\]
Here $E_{\pm}= E \pm\frac{\epsilon }{2}$, $n_{E}^{l}$ and $n_{E}^{r}$
are the electron distribution functions for the energy $E$ in the left
and right regions respectively, and $$T_{E,E'}=
\sum_{p,p'}|V_{p,p'}^{T}|^2 \delta (E - \varepsilon _{p}) \delta (E' -
\varepsilon _{p'}),$$ $\varepsilon_{p}$ being the single particle
energy; one obtains $w_{-}$ by substituting $n^{l,r}$ for $(1-
n^{l,r})$ in $w_{+}$.

The second term, $W_{MP}$, has the form
\[ W_{MP}
= 2i \sin \frac{\lambda}{2}\,\Re\left(e^{i \Lambda } w_{MP}(\tau
)\right)  \; ,
\]
where
\[
w_{MP}(\tau ) = 2\int dE\, d \epsilon \,
T_{E_{-}, E_{+}}
\left( n_{E_{-}}^{l} 
- n_{E_{+}}^{r}\right)\;
\frac{e^{i  \epsilon^{+}\tau}-1}{(\epsilon^{+})^2}
\quad,\quad \epsilon^{+}= \epsilon  +i 0
\; ,
\]
and $W_{MP}$ is thus a $4 \pi -$periodic function of $\lambda $.

The generating function factorizes, $\chi = \chi_{MM} \, \chi_{MP}$,
where $\chi_{MM}= e^{W_{MM}}$ is identical to the result in
Ref.\cite{LevRez01}, and
\begin{equation}
 \chi_{MP}(\tau , \lambda )
=
  \sum\limits_{m= - \infty }^{\infty }
e^{im \lambda }
J_{m}^2\left( w_{MP}(\tau ) \right)\; ,
\label{zvc}
\end{equation} 
$J_{m}$ being the Bessel function.  If $T_{E,E'}$ is featureless on
the scale of the Fermi energy, $E_{F}$, $w_{MP}$ is a constant once
$\tau \gg \hbar/ E_{F} $,
\begin{equation}
w_{MP}(\infty )=  \pi \,\frac{R_{0}}{R_{T}} \; ,
\label{2vc}
\end{equation}
where $R_{T}$ is the tunneling resistance and $R_{0}= 2\pi \hbar/e^2$.

The calculation for the tunnel junction qualitatively agrees with the
result we obtained in the single-particle case.  We observe again that
the formula of Ref.~\cite{LevLeeLes96,LevRez01}, which is identical to
$\chi_{\tau }(\lambda ,\Lambda =0 )$, contains $4 \pi $-periodic
terms.  According to the derivation, they originate from tunneling
events which occur before the measurement started and create charge
off-diagonal elements in the density matrix by the time $t=0$.  For
large enough measuring times, $W_{MP}$ saturates unlike $W_{MM}$ which
grows linearly with time, and $W_{MP}$ represents memory of the
initial state of the system, amounting to intrinsic voltage
independent charge fluctuations. The latter, however, need not be
small, as seen from Eq.~(\ref{2vc}).  According to our analysis, the
results in \cite{LevRez01}, based on Ref.~\cite{LevLeeLes96}, are
valid only if the two electrodes are not connected before the
measurement whereby charge superposition is prevented.

In this paper we have reconsidered counting statistics applying the
rules of quantum mechanics.  We confirm the counting formula of
Levitov {\it et al.}, but only for cases where the initial state of
the system is charge diagonal, {\it i.e.}, when superposition of
different charge states is absent.  Our approach leads to a novel
formula for the probability distribution of integer charge transfers
which is valid for a general charge coherent state.

The role of charge coherence, which is a ubiquitous feature of any
many-body quantum state, should be examined in each particular case.
For a system of {\it non-interacting} particles, one may argue that
details of the initial state are of minor importance since the net
contribution of the non-diagonal elements tends to average out.  Our
tunnel junction results show how it comes about in this particular
case: Even though charge coherence is present in the initial state at
$t=0$, created by prior tunneling events, its contribution expressed
by $w_{MP}$ does not grow with time, diminishing in importance at
large measuring times.  Nevertheless, the charge coherence present at
the start of counting, does noticeably change the statistics in
accordance with our formula Eq.~(\ref{zvc}), especially for short
measuring times.  For {\it interacting} systems the situation is
similar, provided the charge structure of the quantum state can be
expressed in terms of quasi-particles.  An important counterexample is
a superconductor, where the superposition of different charge states
is rigidly maintained, as required by the number-phase uncertainty
relation.  Although the theoretical objects entering the formula of
Levitov {\it et al.}  can be calculated for superconductors
\cite{BelNaz01} and even measured \cite{LevLeeLes96,NazKin}, they
cannot be interpreted as charge transfer probabilities.  Since charge
off-diagonal elements are important, the formula of
Ref.~\cite{LevLeeLes96} cannot be used to calculate the statistics of
charge transfer related to the current of Cooper pairs.  The fact that
the previous counting formula leads to negative ``probabilities'' in
the case of a Josephson junction \cite{BelNaz01}, is understandable in
view of this observation.

In conclusion, we have shown that by using gauge and charge projection
operators to analyze the structure of a quantum state of an arbitrary
system, one is able to construct a probability distribution for charge
transfers of particles obeying quantum dynamics.  The constructed
function is a proper probability distribution, i.e., positive definite
and normalized, and the probability distribution for charge transfer
counting of classical mechanics emerges in the correspondence limit.
The charge transfer in a tunnel junction is considered, and the
modification of counting statistics due to charge coherence has been
demonstrated.

We are grateful to J. Wabnig, Yu. Makhlin, and A. Shnirman for
discussions.  This work was supported by The Swedish Research Council.
The paper was completed during a visit of one of us (A. S.) to
Institut f\"ur Theoretische Festk\"orperphysik, Karlsruhe University,
and the hospitality extended during the visit is greatly appreciated;
financial support from SFB 195 of the DFG is acknowledged.


\begin{thebibliography}{99}

\bibitem[*]{byline}Also at A. F.  Ioffe Physico-Technical

\bibitem{LevLeeLes96}
L. S. Levitov, H. W. Lee, G. B. Lesovik, 
J. of Math. Phys. {\bf 37}, 4845 (1996).

\bibitem{Les89}
G. B. Lesovik, JETP Lett. {\bf 49}, 592 (1989).

\bibitem{Khl87}
V. K. Khlus, Sov. Phys. JETP {\bf 66}, 1243 (1987).

\bibitem{But90}
M. Buttiker, Phys. Rev. Lett. {\bf 65}, 2901 (1990).

\bibitem{BeeBut92}
C. W. J. Beenakker, M. Buttiker, Phys. Rev. B{\bf 46}, 1889 (1992).

\bibitem{BeeSch01}
C. W. J. Beenakker, H. Schomerus,
Phys. Rev. Lett. {\bf 86}, 700 (2001).

\bibitem{BorBelBru02}
   J. B\"orlin, W. Belzig,  C. Bruder,
  Phys. Rev. Lett. {\bf 88}, 197001  (2002);  
P. Samuelsson, M. B\"uttiker, Phys. Rev. Lett. {\bf 89}, 046601  (2002).  

\bibitem{NazKin}
Yu. V. Nazarov, M. Kindermann, cond-mat/0107133.

\bibitem{LevRez01} L. S. Levitov, M. Reznikov, cond-mat/0111057. 

\bibitem{BelNaz01} W. Belzig, Yu. V. Nazarov,
Phys. Rev. Lett. {\bf 87}, 067006 (2001).

 \end{thebibliography}
\end{document}